\documentclass{aa}
\usepackage{graphicx}
\begin{document}
\def\new#1{{#1}}
   \title{A continuous Flaring- to Normal-branch transition in
   \object{Sco~X-1}}

   \author{
           P. Casella\inst{1,}\inst{2},
           T. Belloni\inst{1},
           and L. Stella\inst{2}
          }

   \offprints{casella@merate.mi.astro.it}
   \institute{
              INAF - Osservatorio Astronomico di Brera,
              via E. Bianchi 46, I--23807 Merate (LC), Italy\\
         \and
              INAF - Osservatorio Astronomico di Roma,
              Via di Frascati, 33, I--00040 Monte Porzio Catone (Roma), Italy\\
             }

     \date{Received ..... .., ....; accepted ..... .., ....}

   \abstract{
We report the first resolved rapid transition from a Flaring Branch
Oscillation to a Normal Branch Oscillation in the RXTE data of the Z source
\object{Sco~X-1}. The transition took place on a time scale of $\sim$100
seconds and was clearly associated to the Normal Branch-Flaring Branch vertex
in the color-color diagram. 
We discuss the results in the context of the possible association of the
Normal Branch Oscillation with other oscillations known both in Neutron-Star
and Black-Hole systems, concentrating on the similarities with the narrow 4-6
Hz oscillations observed at high flux in Black-Hole Candidates.
   }
   \titlerunning{Monitoring the FB-NB transition in Sco X-1}
   \authorrunning{P.Casella et al.}
   \maketitle


\section{Introduction}

Neutron star Low Mass X-ray Binaries are usually classified in two
classes, the Z and the atoll sources; these names are based on the shape that
individual sources describe in the color-color diagrams as a result of 
\new{changes taking place in the sources}
(CDs, \cite{HK89}, \cite{vdK05}). The atoll sources
trace a pattern consisting of a single curved branch (banana branch), together
with a smaller ``island'', while the pattern traced by  Z sources is
characterized by three branches called horizontal branch, normal branch and
flaring branch respectively. Three types of \new{Low-frequency Quasi-Periodic
  Oscillations (LFQPOs)} have been associated with the position along the Z:
the HBOs (horizontal branch oscillations), the NBOs (normal branch
oscillations) and the FBOs (flaring branch oscillations).

NBOs with frequencies between 4.5 and 7 Hz have been reported from all Z
sources. Their fractional rms amplitude is typically between 1 and 3 \% in the
1-10 keV band, and are strongest near the middle of the NB. FBOs were seen
only in two sources (\object{Sco~X-1}, \cite{Priedhorsky86};
\object{GX~17+2}, \cite{Homanetal02}), while a broad excess that
may be due to an FBO peak moving rapidly in frequency as the source moves
rapidly up and down the FB has been reported in other sources (see
e.g. \object{GX~5-1}, \cite{Jonkeretal02}). FBOs occur on a small part
($\sim$10 \% of the total
extent) of the Flaring Branch (FB) nearest the NB, and have frequencies in
the range $\sim$6-25 Hz, increasing from the NB-FB vertex through the FB. With
increasing frequency, the fractional amplitude of the FBOs remains
approximately constant, while its width increases until the peak becomes too
broad to distinguish from the broad band noise in the \new{power density
spectrum} (the so-called high frequency noise, HFN).

NBOs and FBOs are thought to be physically related,
since in \object{Sco~X-1} the NBO frequency joins to the FBO frequency as the
source moves from the NB to the FB. However, the increase from $\sim$6 Hz to
$\sim$10 Hz occurs in a very short segment of the Z track located just at the
vertex of the NB-FB junction, and is at present still unresolved
(\cite{dieters&vdk00}). The transition is even less clear in \object{GX~17+2}:
\cite{Homanetal02}) found that near the NB-FB vertex the frequency of the FBO was a
factor of $\sim$2 higher than that of the NBO, which could mean that the NBO
and FBO are harmonically related. However, by inspecting all single dynamical
power spectra of observations with values of the curvilinear coordinate $S_z$
(tracking the position along the Z pattern, see \cite{vdK95}) around 2
(corresponding to the NB-FB vertex), they found in some cases QPOs with
intermediate frequencies ($\sim$10 Hz), suggesting that the frequency does not
jump directly from $\sim$7 to $\sim$14 Hz. However no clear transitions were
found.

At present, it is currently believed that NBOs and FBOs are different
manifestations of the same phenomenon, the properties of which rapidly change
\new{at around the position} of the lower vertex in the Z track. However, the
transition between them has not yet been unambiguously resolved. In this paper
we present the analysis of a RXTE observation of \object{Sco~X-1} in which for
the first time a clear continuous fast transition from a Normal Branch
Oscillation to a Flaring Branch Oscillation is seen and which is associated
with the NB-FB vertex in the Z track.


\section{Data analysis}

We analyzed a RXTE/PCA observation of \object{Sco~X-1}, from the RXTE public
archive, made on MJD 50230 (1996-05-27). The observation consists of three
continuous data intervals of $\sim$1600, $\sim$550 and $\sim$2000 s duration
respectively. All five PCA units were on during the whole observation. The PCA
data were obtained in several simultaneous different modes (see
Tab. \ref{XTEmodes}). During the first interval, the pointing direction was
offset by $\sim$0.3 deg, while in the other two intervals the offset was
reduced to $\sim$0.005 deg. For this reason we could not build the expected
Z-track across the whole observation. We concentrated our analysis on the first
of the three data intervals, during which a variable low frequency ($\sim$6-15
Hz) QPO was detected. 


   \begin{table}
\scriptsize
      \caption[]{RXTE/PCA data modes active during the \object{Sco~X-1}
      observation}
         \label{XTEmodes}
     $$
         \begin{array}{llcc}
            \hline
            \noalign{\smallskip}
            Mode & Time~res. & Number~of & PHA \\
            Name & \hspace{2.0mm}(s.) & PHA~Channels & Energy~range~(keV) \\
            \noalign{\smallskip}
            \hline
            \noalign{\smallskip}
            {\tt Standard1} & \hspace{4.0mm}2^{-3}  & 1   & 2-60  \\
            {\tt Standard2} & \hspace{4.0mm}2^{~4}  & 128 & 2-60  \\
            {\tt Binned1}   & \hspace{4.0mm}2^{-13} & 1   & 2-32  \\
	    {\tt Binned2}   & \hspace{4.0mm}2^{-8}  & 16  & 2-60  \\
            {\tt SB}        & \hspace{4.0mm}2^{-13} & 1   & 2-60  \\
            \noalign{\smallskip}
            \hline
         \end{array}
     $$
   \end{table}

\begin{figure}[t]
     \centering
     \includegraphics[width=8.5cm]{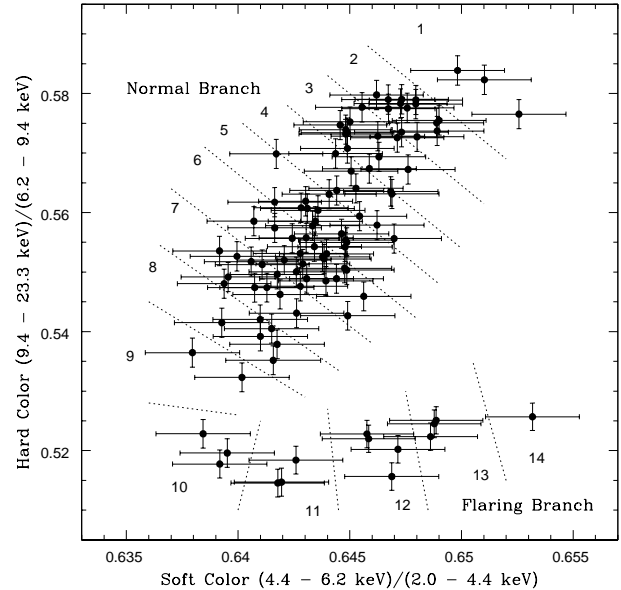}
     \caption{Color-color diagram of the first $\sim$1600 seconds. Time
     resolution is 16 s. Dotted lines indicate the intervals selection for the
     timing analysis.}
     \label{ccSco}
   \end{figure}

We used {\tt Standard2} data to produce a
color-color diagram (CD). For each 16 s data segment (the intrinsic
resolution of the {\tt Standard2} mode), we defined two colors as
ratios of count rates in two different energy bands. The energy bands used for
the colors (soft and hard color) are given in Table \ref{scocolors}. The CD is
shown in Figure \ref{ccSco}: \object{Sco~X-1} was in its Flaring Branch (FB)
at the beginning of the observation, and moved into the Normal Branch (NB)
after $\sim$200 seconds.

\begin{table}[h]
\scriptsize
\caption{Channel and energy boundaries of the soft and hard colors}
\label{scocolors}
$$
\begin{array}{lcc}
\hline
\noalign{\smallskip}
Color & Channels & Energ (keV) \\
\noalign{\smallskip}
\hline
\noalign{\smallskip}
$Soft$ & 12-16/0-11 & 4.4-6.2/2.0-4.4 \\
$Hard$ & 26-63/17-25  & 9.4-23.3/6.2-9.4 \\
\noalign{\smallskip}
\hline
\end{array}
$$
\end{table}

\begin{figure}[t]
     \centering
     \includegraphics[width=8.5cm]{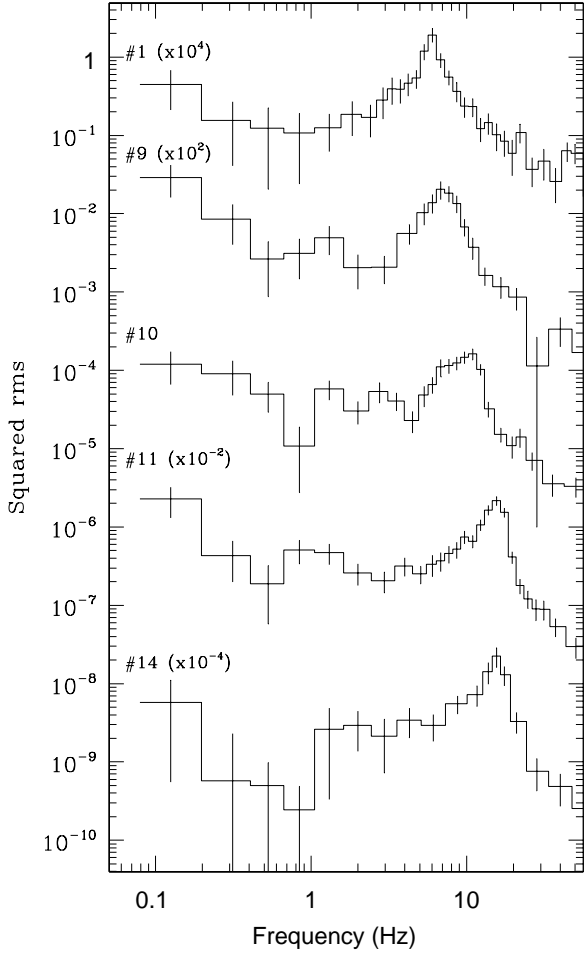}
     \caption{Five power spectra corresponding to the extremes (\#1 and \#14)
     and to the vertex (\#9, \#10 and \#11) of the CD. Spectra are shifted by
     the indicated factors for clarity.}
     \label{sco5pds}
   \end{figure}

Since the Z track is not complete, we could not use the usual rank
number $S_z$ used in literature, which is defined using both vertices
(Hasinger et al. 1990). In order to check the presence of a QPO along the Z
track, we divided the CD in 14 intervals and calculated a power spectrum
(with a Nyquist frequency of 128 Hz) for each of them by averaging power
spectra calculated every 16 s data interval.
The two power spectra corresponding to the extremes of the CD (\#1 and \#14),
together with three power spectra \new{approximately} corresponding to the
vertex (\#9, \#10 and \#11) are shown in Figure \ref{sco5pds}. 

We then fitted each power spectrum with a combination of a simple power law and
 a broad Lorentzian shape, approximating the broad band noise, and a
 narrow Lorentzian shape, approximating the QPO peak. For the
 intervals in the FB an additional narrow ($\Delta\nu/\nu\ga$1) Lorentzian
 shape was needed, in order to better approximate the asymmetric profile of
 the FBO peak. \new{PDS fitting was carried out with the
standard Xspec fitting package, by using a one-to-one energy-frequency
conversion and a unit response}.
Because of the low intensity of the broad band noise, the parameters
of the power law and those of the broad Lorentzian were not
well constrained, and their behavior could not be studied. However, since the
intensity of the QPO peak was high in all intervals, the uncertainties in the
underlying continuum did not impact much on the estimate of the peak
parameters.

   \begin{figure}[t]
     \centering
     \includegraphics[width=8.5cm]{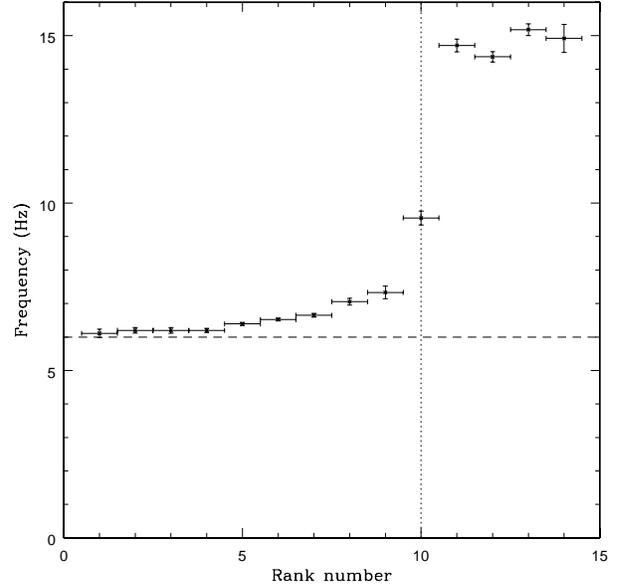}
     \includegraphics[width=8.5cm]{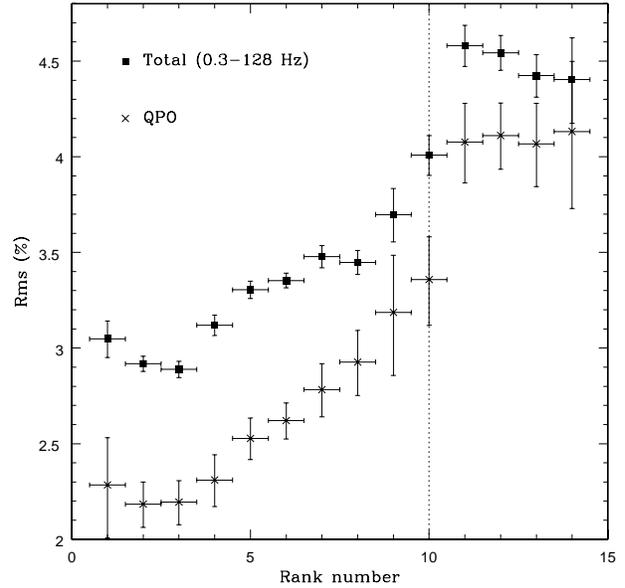}
     \caption{QPO centroid frequency ({\it upper panel}) and fractional rms of
     the QPO and of the whole power spectrum ({\it bottom panel}) as a
     function of the interval number through the color-color diagram. The
     \new{h}orizontal line in the upper panel indicates the ``6 Hz
     asymptote''. The vertical line in both panels \new{corresponds to} the
     vertex in the Z track (see Figure \ref{ccSco})
     }
     \label{scoRmurms}
   \end{figure}

In the upper panel of Figure \ref{scoRmurms} we plot the QPO centroid frequency
as a function of the interval number. The evolution of the QPO peak is
evident: the centroid frequency was $\sim$6 Hz when the source was in the upper
part of the visible NB, slowly increased up to $\sim$7.5 Hz when the source
approached the FB-NB vertex, then quickly increased up to $\sim$15 Hz in
the FB. In the bottom panel of Figure \ref{scoRmurms}, the total
fractional rms and the QPO fractional rms are shown as a function of the
interval number.
Both slowly increase through the NB towards the vertex, and reach
their maximum in the FB. The total rms slightly decreases after the vertex,
while the QPO rms is consistent with remaining constant.

\begin{figure}[t]
     \centering
     \includegraphics[width=8.5cm]{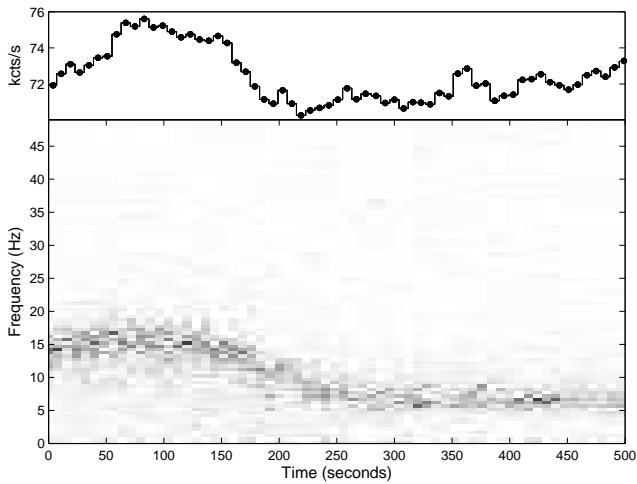}
     \caption{2-13 keV light curve ({\it upper panel}) and dynamical power
     spectrum ({\it bottom panel}) of the first 500 seconds of the
     observation.}
     \label{scoPDS}
   \end{figure}

The transition appears to be very fast, as a result of the fast
passage of the source through the NB-FB vertex. 
In order to track more finely the QPO frequency as a function of time, we
produced a dynamical power spectrum for the
whole observation by calculating a power spectrum every 4 seconds over PCA
channels 0-35 (corresponding approximately to the 2-13 keV energy range) and
with a time resolution of 1/128 s (corresponding to a Nyquist frequency
of 64 Hz). The dynamical power spectrum and the correspondent light curve of
the first 500 seconds of the observation, when the frequency transition
occurred, are shown in Figure \ref{scoPDS}. 

The transition is clearly visible in the dynamical power spectrum: the QPO
peak at the beginning of the observation (when the source is in its FB) is at
frequencies $\sim$14 Hz, has a maximum at $\sim$15 Hz after \new{less than}
100 seconds, and rapidly decreases (corresponding to the FB-NB vertex) after
few hundred seconds to $\sim$6 Hz. The peak remains then visible in the
dynamical power spectrum at $\sim$6 Hz in the remaining part of the
observation, during which the source is in the NB. 

\section{Discussion}

We presented the first continuous monitoring of \new{a rapid transition between
Flaring Branch and Normal Branch Oscillations}. Even though the transition
could not be \new{continuously tracked through} the Z pattern in the CD, the
high statistics of the data makes it visible in the dynamical power
spectrum, where it is resolved to take place in $\sim$100 seconds. 

The presence of the NBO/FBO in the Z sources has often been related to the
fact that these sources accrete at near-Eddington mass accretion rates. It is
thought that at these high mass accretion rates a significant fraction of the
accretion flow is in the form of a thick, perhaps nearly spherical flow. Also,
the
effects of radiation pressure are thought to play an important role in this
regime. Fortner, Lamb \& Miller (1989) suggested that the NBOs are the result
of a radiation force/opacity feed-back mechanism within a spherical flow
region. For reasonable physical parameters, the frequency of the NBOs is
predicted to be $\sim$6 Hz at the onset of the process, and then to increase
to $\sim$10 Hz as
the luminosity approaches $L_{Edd}$. These oscillations can furthermore excite
other modes with {\it similar} frequencies, which are expected to increase as
the luminosity approaches and then exceeds $L_{Edd}$. However, it is not clear
if the similarity of these two oscillation types can be as high as required to
explain the smoothness of the transition reported in \object{Sco~X-1} (see
Figure \ref{scoPDS}). Another model for the NBOs was proposed by Alpar et
al. (1992), in which the NBO frequency is basically that of sound waves in a
thickened accretion disk. However, this model does not explain how the NBO
changes into the FBO. A strong challenge for these models is the recent
discovery of $\sim$6 Hz QPOs, with properties similar to those of NBOs, in
atoll sources (\new{see e.g. Wijnands, van der Klis \& Rijkhorst 1999,
Wijnands \& van der Klis 1999,} Belloni, Parolin \& Casella 2004), in which
the mass accretion rate is always lower than the Eddington limit.

\new{Casella, Belloni \& Stella (2005) recently reviewed and further extended
  the similarities between the LFQPOs observed in Z sources and the three main
  types of LFQPOs observed in Black-Hole Candidates (BHCs, see the cited work
  and references therein), and proposed a one-to-one association where the C-,
  B- and A-Type observed in BHCs correspond respectively to the HBO, NBO and
  FBO observed in Z sources. Whatever the physical mechanisms that determines
  these oscillations are, the presence of such mechanisms in both types of
compact objects would clearly favor a disk origin for these
oscillations, ruling out all models that involve any interaction with
the surface or the magnetosphere of the neutron star (see e.g. Stella, Vietri
\& Morsink 1999; see also Kluzniak 2004).
}

In this context, the smooth continuous transition between
the NBOs and the FBOs appears to be very different from the fast, unresolved
transitions observed between type-B and type-A QPOs in BHCs (Nespoli et
al. 2003; Casella et al. 2004), i.e. the
equivalent of NBOs and FBOs in this analogy. At present it is in fact not clear
whether type-B and type-A QPOs originate from the same physical phenomenon,
the properties of which
can change in a short time scale (less than a few tens of seconds in the known
cases), or if they are two distinct
phenomena, one of which (type-A) is too weak to be detected when the other
(type-B) is present (see e.g. Nespoli et al. 2003).
\new{Available data at present do not permit us to address this issue,
and more observations (and/or better statistics) are needed.
}


\begin{acknowledgements}
This work was partially supported by the Italian Ministry of Education,
University and Research under CO-FIN grants 2002027145 and 2003027534.
\end{acknowledgements}




\begin{thebibliography}{}

\bibitem[Alpar \& Shaham 1985]{Alpar&Shaham85}
Alpar, M. A., \& Shaham, J., 1985, Nature, 316, 239

\bibitem[Alpar et al. 1992]{Aplaretal92}
Alpar, M. A., Hasinger, G., Shaham, J., \& Yancopoulos, S., 1992, A\&A, 257,
627

\bibitem[Belloni, Parolin \& Casella 2004]{1820}
Belloni, T., Parolin, I., \& Casella, P., 2004, A\&A, 423, 969

\bibitem[Casella et al. 2004]{Casellaetal04}
Casella, P., Belloni, T., Stella, L., \& Homan, J., 2004, A\&A, 426, 587

\bibitem[Casella, Belloni \& Stella 2005]{CBS05}
Casella, P., Belloni, T., \& Stella, L., 2005, ApJ, accepted (preprint:
astro-ph/0504318)

\bibitem[Dieters \& van der Klis 2000]{dieters&vdk00}
Dieters, S. W., \& van der Klis, M., 2000, MNRAS, 311, 201

\bibitem[Fortner, Lamb \& Miller 1989]{fl&m89}
Fortner, B., Lamb, F. K., \& Miller, G. S, 1989, Nature, 342, 775

\bibitem[Hasinger \& van der Klis 1989]{HK89}
Hasinger, G., \& van der Klis, M., 1989, A\&A, 225, 79

\bibitem[Homan et al. 2002]{Homanetal02}
Homan, J., van der Klis, M., Jonker, P. G., Wijnands, R., Kuulkers, E.,
Mendez, M., \& Lewin, W. H. G., 2002, ApJ, 568, 878

\bibitem[Jonker et al. 2002]{Jonkeretal02}
Jonker, P. G., van der Klis, M., Homan, J., Mendez, M., Lewin, W. H. G.,
Wijnands, R., \& Zhang, W., 2002, MNRAS, 333, 665

\bibitem[Kluzniak 2004]{Kluzniak04}
Kluszniak, W., 2004, Revista Mexicana de Astronomía y Astrofísica (Serie de Conferencias) Vol. 20. IAU Colloquium 194, pp. 128-129 (2004)

\bibitem[Nespoli et al. 2003]{Nespolietal03}
Nespoli, E., Belloni, T., Homan, J., Miller, J. M., Lewin, W. H. G., Mendez,
M., \& van der Klis, M., 2003, A\&A, 412, 235

\bibitem[Priedhorsky et al. 1986]{Priedhorsky86}
Priedhorsky, W., Hasinger, G., Lewin, W. H. G., Middleditch, J., Parmar, A.,
Stella, L., \& White, N., 1986, ApJ, 306, L91

\bibitem[Stella, Vietri \& Morsink 1999]{Stella99}
Stella, L., Vietri, M., Morsink, S. M., 1999, ApJ, 524, L63

\bibitem[van der Klis 1995]{vdK95}
van der Klis, M., 1995, in ``X-ray binaries'', eds. Lewin, W. H. G., van
Paradijs, J. and van den Heuvel, E. P. J., Cambridge University Press,
Cambridge, p. 252

\bibitem[van der Klis 2005]{vdK05}
van der Klis, M., 2005 in ``Compact Stellar X-Ray Sources'',
eds. W.H.G. Lewin and M. van der Klis, Cambridge University Press, in press

\bibitem[Wijnands \& van der Klis 1999]{W&vdK99}
Wijnands, R., \& van der Klis, M., 1999, ApJ, 522, 965

\bibitem[Wijnands, van der Klis \& Rijkhorst 1999]{Wijnandsetal99}
Wijnands, R., van der Klis, M., \& Rijkhorst, E., 1999, ApJ, 512, L39

\end{thebibliography}
\end{document}